# Concentrator of elastic waves

*S. N. Dolya*

*Joint Institute for Nuclear Research, Joliot Curie str. 6, Dubna, Russia, 141980*

This article is dedicated to an opportunity of concentrating elastic waves in the iron and water cones on the square of the cone vertex of the order of 1 cm$^2$. The square of the base of the cone is equal to 1 m$^2$, its height - 1 m. The calculations assume that the cone hexogen network lying in the cone basis explodes during the time of 1 μs and causes an explosive wave converging to the vertex of the cone. It is shown that this explosive wave can accelerate the body having a mass of 3 g to speed V = 5 km / s.

## 1. Physical motivation of the task

Suppose there is an iron cone, which is a part of a ball with radius $r_0 = 1$ m and a basic square of $S_{cone} = 1$ m$^2$. On a spherical base of the cone let us place a grid with a cell size h = 3 cm, in whose nodes there are hexogen balls with a diameter of $d_{hex} = 1$ cm. The volume of the ball $V_{hex} = \pi d^3_{hex} / 6 \approx d^3_{hex} / 2$ or 0.5 cm$^3$. At a density of RDX [1] $\rho_{hex} \approx 1.8$ g / cm$^3$, the mass of each ball is about 1g. Obviously, that the total number of balls placed on the base of the cone is 10$^3$, and their total mass $m_{hex} = 1$ kg.

At the simultaneous undermining of all the balls, the released energy will be of the order of $\Delta Q_{hex} \approx 6$ MJ. The velocity of the detonation wave in RDX is $V_{det.\ wave} = 8.36$ km / s, so that we can assume that the time of the energy release will be $\Delta \tau_{hex} = 1$ μs. The power of the energy release will be equal to the following:

$$W_{hex} = \frac{1}{2} \Delta Q_{hex} / \Delta \tau_{hex} = 3 \ast 10^{12}\ W, \qquad (1)$$

where we have taken into account that only half of the energy released in the explosion will get inside the cone.

The density of energy release power that is the intensity of the wave (pulse) near the base of the cone is equal to:

$$I_1 = W_{hex} / S_{cone} = 3 \ast 10^{12}\ W/m^2. \qquad (2)$$

In the region with an area of $S_{vert} = 1$ cm² at the cone vertex there will be concentration of the explosion energy, the intensity of the pulse there will be bigger if to take the square of the base and the square of the vertex:

$$I_2 = I_1\, S_{cone}/\, S_{vert} = 3*10^{16}\ W/m^2. \qquad (3)$$

**2. The amplitude of oscillations**

The pressure in the sonic wave is related with the sound intensity by the following ratio:

$$P_{sound\ Fe} = (I_2 * \rho_{Fe} * V_{sound\ Fe})^{1/2}, \qquad (4)$$

where $I_2$ – the sound intensity near the vertex, $\rho_{Fe} = 9 * 10^3$ kg / m³ – the density of iron, $V_{sound\ Fe} \approx 6$ km / s – the velocity of longitudinal elastic waves in iron [2], p. 86.

Substituting numbers into the formula (4), we find that the pulse pressure near the vertex of the cone is equal to: $P_{Fe} = 1.3 * 10^{12}$ Pa.

Now we find the rate of displacement of atoms $V_{disp}$ in this pulse from the following relation:

$$P_{Fe} = \rho_{Fe} * V^2_{disp}/2, \qquad (5)$$

where

$$V_{disp} = (2 * P_{Fe}/\, \rho_{Fe})^{1/2} = 1.7*10^4\ m/s. \qquad (6)$$

We have obtained the velocity of the atom displacements to be equal to $V_{disp} = 17$ km / s, which is by twice larger than the first space velocity $V_{1\ space} = 8$ km /s.

This shows that if to place a small body in the vertex of such a cone, a significant portion of the explosion energy will be transferred to this body which can be launched into space.

It can be seen that the iron cone will be damaged with each explosion and its reuse is hardly possible.

## 3. Water cone

Let us consider an opportunity of concentrating the energy of elastic oscillations by using the water cone having the same sizes as the iron one. Water may be placed into the Mylar or rubber shell. The volume of the cone is $V_{cone} = (1/3)\, r^3_0 = 0.3$ m$^3$. When the density of water $\rho_{water} = 10^3$ kg / m$^3$, the mass of water in this cone, will be approximately equal to $m_{water} = 300$ kg. The cone should be positioned inside a hard steel trunk to prevent the damage of the water cone in the transverse direction.

We find the pressure in the sonic wave near the cone vertex:

$$P_{sound\ water} = (I_2 * \rho_{water} * V_{sound\ water})^{1/2}. \qquad (7)$$

Substituting numbers into the formula (7), where $\rho_{water} = 10^3$ kg / m$^3$, $V_{sound\ water} = 1.5$ km / s, we find that $P_{sound\ water} = 2 * 10^{11}$ Pa.

The rate of displacement of water molecules in the pulse is equal to:

$$V_{disp} = (2 * P_{water} / \rho_{water})^{1/2} = 1.4 * 10^4 \text{ m/s}, \qquad (9)$$

it also turned out to be greater than the first space velocity.

## 4. Non-linear theory

It is clear that at such high pressures of $P_{sound\ water} = 2 * 10^{11}$ Pa, the density of water will change, and together with it the other parameters will also change. The system of equations describing the state of condensed water can be represented as follows [3]:

$$\rho/\rho_0 = V_s/(V_s - V_d); \qquad (10)$$

$$P_s + \rho(V_s - V_d)^2 = \rho V_s^2; \qquad (11)$$

$$P_s = A[(\rho/\rho_0)^n - 1]; \qquad (12)$$

where $\rho_0$ and $\rho$ are the density of water in front and behind the explosive wave front, $V_s$ – the velocity of propagation of the explosive wave front, $V_d$ - the mass velocity behind the explosive wave front, $A = 3 * 10^8$ Pa, $n = 7 - 8$. Equation (10) expresses the law of conservation of mass, equation (11) - the law of conservation of momentum, and the equation (12) is the equation of state of water in the compressed form.

Let $\rho / \rho_0 = 3$, $V_s = 15$ km / s. Then the pressure in the sound wave will be more than the linear pressure (equation (7)) by $(30)^{1/2}$ times, i.e. it is equal to $P_s \approx 10^{12}$ Pa. From (12) we find that $P_s$ for $\rho / \rho_0 = 3$ and $n = 8$ is equal to $P_s = 2 * 10^{12}$, which can be considered to be a good agreement between the above.

From equation (10) we find that in this case that the mass velocity behind the explosive wave front will be equal to $V_d = 10$ km / s.

We substitute these values in the formula (11). We find:

$$10^{12} + 7.5*10^{10} = 6.75*10^{11}.$$

That may be also considered a rather good agreement.

**5. Energy transfer to the body**

Suppose that in the vicinity of the cone vertex there is is a physical body with a cross-section of 1 cm$^2$ and a mass of $m_b = 3$g. From the law of conservation of momentum it follows that in this case the velocity of $V_b = 5$ km / s will be transferred to the body.

The kinetic energy of the body in this case will be equal to:

$$m_b V_b^2 / 2 = 37.5 \text{ kJ}, \qquad (13)$$

and the coefficient of energy transfer from the explosion to the physical body will be equal to 37 kJ / 6 MJ ≈ 0.6%.

Let us consider the maximum lifting height and range of the flight of the body, released at the angle of $45^0$ to the horizon, taking into account the air resistance. The velocity of the body due to this air resistance decreases with time of flight according to the law [4]:

$$V(t) = V_0/(1+\rho_0*\exp(-z/H_0)*C_x*S_{tr}* V_0 t/2m), \qquad (14)$$

where $V_0$ - initial velocity, $\rho_0 = 1.3 * 10^{-3}$ g / cm$^3$ - the density of the air near the Earth surface, $H_0 = 7$ km – a barometric coefficient, m - mass of the body, $C_x$ - drag coefficient of the body, $S_{tr}$ - cross-section of the body, z - lifting height of the body.

Without the air resistance the formulae of the flight range and maximum lifting height are as follows:

$$S_{max} = 2V^2_0 * \sin\Theta * \cos\Theta / g, \qquad (15)$$

$$H_{max} = V^2_0 * \sin\Theta * \cos\Theta / 2g, \qquad (16)$$

where $g = 10^{-2}$ km / s$^2$. Then for $V_0 = 5$ km / s and the angle of inclination to the horizontal velocity of $\Theta = 45^0$, we would have obtained the maximum flight range $S_{max} = 2500$ km and maximum lifting height $H_{max} = 625$ km.

However, due to the air resistance, the velocity V (t) decreases according to the law given by the formula (14). The density of the atmosphere decreases exponentially according to the barometric formula. At the height of H = 7 km it is already less by e times than that at the Earth surface, where e = 2.72 – the base of natural logarithms.

If you use a bullet, whose diameter is much less than the diameter of the trunk, then for this bullet it is possible to have a small drag coefficient $C_x$.

The reason why we can make a small drag coefficient lies in its dependence on the velocity for a sharp cone. At hypersonic velocities the drug coefficient does not depend on the velocity of the body and becomes equal to a constant value:

$$C_x = \Theta^2_{vert}, \qquad (17)$$

where $\Theta_{vert}$ - the angle at the vertex of the cone. Fo the body of an elongated form, which has the ratio of the body length $l_{body}$ to its diameter $d_{body}$ much greater than the unity $l_{body} / d_{body} >> 1$, the angle at the vertex of the vertex can be made sufficiently small, for example, $\Theta_{vert} = d_{body} / l_{body} = 0.1$. Then the drag coefficient $C_x = 10^{-2}$, and it can be expected that the decrease of the body velocity while it's crossing the atmosphere will be small. After crossing the atmosphere by the body you can use formulae (15) and (16) by substituting the appropriate velocity value into the formulae, which is obtained after the body crossing of the atmosphere.

Here it is necessary to take into account that the body rises into the atmosphere not vertically, but at the angle of $45^0$. This can be taken into account by dividing the second term in the denominator by sin Θ, i.e., multiplying it in this case by 1.41.

For the body with a diameter $d_{body} = 3$ mm, the vertex angle $Θ_{vert} = 0.1$, the mass $m_{body} = 3$ g, after three seconds of the flight we obtain: V = 3.4 km / s. The average velocity of the body at this distance will be equal to $V^- = 4.2$ km / s, the lifting height during this lifting time will be 9 km. We assume that further the body will fly without air resistance. Then, according to (15), (16) the body with velocity V = 3.4 km / s will fly $S_{max1} = 1100$ km and rise to the height of $H_{max1} = 290$ km.

**Conclusion**

It is clear that though this concentrator of elastic waves allows it to transfer the hypersonic velocity to a small body, but the efficiency of energy transfer is small.

References

1. https://ru.wikipedia.org/wiki/Гексоген

2. Tables of physical quantities. Handbook ed. I. K. Kikoin, Moscow, Atomizdat, 1976

3. A. G. Russian, V. I. Oreshkin, A. Yu. Lyubetskii et al. Study electrical explosion of conductors in the high pressure of the converging shock wave, Zh. Tech. Phys, t.77, issue 5, p. 35, 2007, http://journals.ioffe.ru/jtf/2007/05/p35-40.pdf

4. http://arxiv.org/ftp/arxiv/papers/1403/1403.4541.pdf